\documentclass[smallextended]{svjour3}       
\smartqed  
\usepackage{graphicx}
\usepackage{amsmath,mathrsfs,graphicx,amssymb}
\newcommand{\ba}{\begin{eqnarray}}
\newcommand{\ea}{\end{eqnarray}}
\renewcommand{\rho}{\varrho}

\def\pt{{\partial}}
\def\ev{{\bf e}}
\def\s0{{s_0}}

\def\nv{{\bf n}}

\def\tv{{\bf t}}

\def\rv{{\bf r}}

\def\tv{{\bf t}}

\def\F{{\rm F}}
\def\E{{\rm E}}

\def\bS{{\bar{S}}}
\def\bx{{\bar{x}}}
\def\lb{{\bar \lambda}}

\newcommand{\veps}{\varepsilon}

\def\t0{{\theta_0}}

\def\de{\delta}

\def\eps{\epsilon}
\def\d{{\rm d}}

\newcommand{\Wb}{W_{\text{buck}}}
\newcommand{\Wa}{W_{\text{adh}}}

\begin{document}

\title{The delamination of a growing elastic sheet with adhesion
}


\author{Gaetano Napoli         \and
        Stefano Turzi 
}


\institute{G. Napoli \at
              Dipartimento di Ingegneria dell'Innovazione, Universit\`a del Salento \\
              Tel.: +39-0832-297799\\
              \email{gaetano.napoli@unisalento.it}           
           \and
           S. Turzi \at
           Dipartimento di Matematica, Politecnico di Milano \\
           Tel.: +39-02-23994650\\
           \email{stefano.turzi@polimi.it}           
}

\date{Received: date / Accepted: date}

\maketitle

\begin{abstract}
We study the onset of delamination blisters in a growing elastic sheet adhered to a flat stiff substrate. When the ends of the sheet are kept fixed, its growth arouses residual stresses that lead to delamination.  This instability can be viewed as a discontinuous buckling  between the complete adhered solution and the buckled solution.  We provide an analytic expression for the critical deformation at which the instability occurs. We show that the critical threshold scales with a single dimensionless parameter that comprises information from the geometry of the sheet, the mechanical parameters of material and the adhesive features of the substrate.

 \PACS{PACS code1 \and PACS code2 \and more}
\end{abstract}

\section{Introduction}
Morphological instabilities and surface wrinkling of soft materials such as gels, elastic films and biological tissues are of growing interest to a number of academic disciplines including the design of technological devices and biomedical engineering. In a broad range of technological applications (for instance, protective coatings, multilayer capacitors, adhesive joints, and stretchable electronic devices) flexible thin films adhere to stiff substrates.  These films, subject to an in-plane compression, are susceptible to buckling phenomena such as delamination: localised regions of the sheet where the film and the substrate are no longer bonded. These delamination blisters are not only interesting in technology  but of eminent importance for biology. Structures reminiscent of the blisters can be found in the development of organs \cite{goriely:2015,ciarletta:2012} as well as in bacteria biofilms \cite{wilking:2013,benamar:2014}.

Blisters in a thin structure can form essentially for two reasons: (i) the two far edges of the strip are brought closer \cite{Wagner:2013,stoop:2015} or (ii) there exists an incompatibility between the geometries of the sheet and substrate. Such a mismatch can be induced, for instance, by thermal dilatation or by growth. In \cite{Vella:2009}, the authors determine the critical threshold for the blistering of an adhesive film coating an elastic substrate. The critical threshold depends on the balance between the elastic and adhesive energies of the substrate  and the bending energy of the film. The birth of a blister related to the growth or compression of a sheet inside a curved cylinder has been analysed in a previous paper \cite{napoli:2015}.  Lee and {\it al.} \cite{lee:2016} studied the onset of a ruck in an heavy carpet. In both latter problems  the finite compressibility of the sheet is the key assumption to study the birth of the instability: blistering (or rucking) can occur at finite end-end 
compression, and with finite compressive load.  On the contrary, the classical analysis, where the sheet is assumed to be unstretchable, leads to the paradox that the compressive force needed to create a blister should be infinite.

Here we consider the case of an elastic film adhering to a flat substrate. We suppose that the film undergoes a homogeneous growth within a rigid container of fixed length $2a$. The growth is modelled by a variable natural length $L$, {\it i.e.} the strip length in absence of external loads. Generally, such a growth is a very slow process compared with the intrinsic dynamics of sheet. This allows us to consider $L$ as an adjustable parameter and to use the equilibrium equations to determine the shape of sheet.     

The paper is organized as follows. In Section \ref{sec:1}, we present a variational derivation of the equilibrium equations, subject to suitable geometrical constraints. The variational procedure is here slightly complicated by the fact that the end-points of the energy functional are not fixed, but are part of the unknowns.  In Section \ref{sec:2}, we perform a numerical simulation of the equilibrium stored energy for different solutions. The  buckled solution exhibits a bifurcation point (a cusp) and two branches. We prove that for small growths, the completely adhered solution is energetically favoured; by contrast, the delaminated solution attains the energy minimum beyond a critical excess-length. In Section  \ref{sec:3}, we  provide the asymptotic approximation of this critical threshold. In Section \ref{sec:4}, we reach the conclusion and add some final comments. 

\label{intro}
 
\section{Variational formulation}
\label{sec:1}
\begin{figure}
\centerline{\includegraphics[scale=1.2]{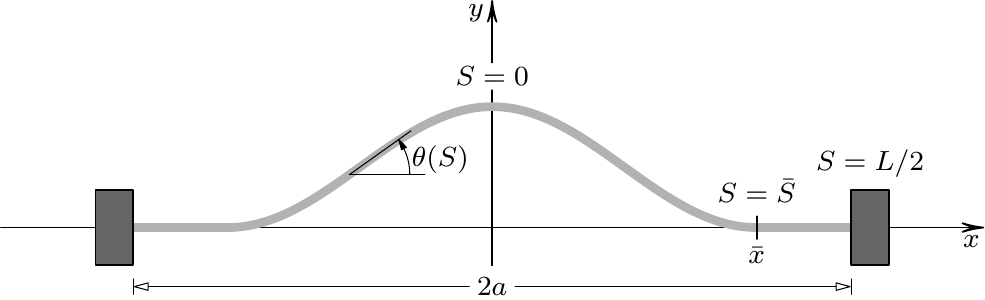} }
\caption{\label{fig:geometria} Schematic representation of the sheet deformation. The strip has a reference length $L$ and is limited by two blocks placed at a fixed distance $2a$.}
\end{figure}

In this Section we derive the equilibrium equations and the boundary conditions as stationary points of the energy functional. We assume that there is no deformation in $z$ direction, so that the longitudinal profile of the sheet can be regarded as a stretchable and flexible rod belonging to the $(x,y)$-plane. This is represented by a parametric curve $\rv(S)$ (see Figure \ref{fig:geometria}), with $S \in [-L/2, L/2]$, where $L$ denotes the length of the strip in the stress-free configuration and $S$ is the referential arc-length. We denote with $\bS$ and $\bx$ the arc-length and the abscissa of the detachment point, respectively.

In the plane of the curve, we introduce a Cartesian frame of reference $(O;\ev_x,\ev_y)$, where $O$ is the origin and $\ev_x$, $\ev_y$ are the unit vectors along, respectively, the $x$ and the $y$ axes.  We  parametrize the tangent and the normal unit vectors by
\[
\tv(S) = \cos \theta(S) \ev_x + \sin \theta(S) \ev_y, \qquad \nv(S) = -\sin \theta(S) \ev_x + \cos \theta(S) \ev_y
\]
and, hence, 
$
\ev_z = \tv \times \nv.
$

The end points of the curve are placed at $(-a,0)$ and $(a,0)$. The energy functional consists of two contributions: the energy of the buckled region, $W_f$, and the energy of the adhered region, $W_a$. Furthermore, the effective energy of the detached part of the beam comprises the bending energy and the compression energy:
\ba
W_f = \int_0^\bS [ k \; \theta'^2 + b (\lambda -1)^2 ]\d S.
\label{libera}
\ea
The constants $k$ and $b$ are positive parameters that represent the bending and the stretching rigidity, respectively.  The intrinsic characteristic length 
\[
\ell := \sqrt\frac{k}{b}
\]
is of the order of the sheet thickness. 

In the adhered region, the sheet is in contact with the flat substrate  and, therefore, no  bending energy is expended by the sheet. However, it may undergo a stretch and, hence, there is an energy cost associated with the stretching term. In addition, we have to consider the contribution of the sheet-substrate adhesion energy. This latter term penalises the delamination and it is assumed to be proportional to the length of the adhered curve. Thus, the total energy of the adhered region is
\ba
W_a = \int_{\bS}^{L/2} \left[ b (\lambda -1)^2 - 2 w \lambda \right]\d S,
\ea  
where $w$ is a positive constant that represents the interfacial energy density. In this paper, we are particularly concerned with the case $w \ll b$. The presence of adhesion introduces a further characteristic length, namely the {\it elasto-capillary} length of the system, that relates the bending and the adhesion terms
\ba
\ell_{ec} := \sqrt\frac{k}{w}.
\ea
A Lagrange multiplier $\mu$ is then used to account for the geometrical constraint that the distance of end-points of the curve on the $x$-axis is fixed. Hence, the effective total energy takes the form
\ba
W = W_f + W_a + 2 \mu\left(a - \int_{0}^{\bS} \lambda \cos \theta \d S - \int_{\bS}^{L/2} \lambda \d S \right).
\label{eq:effective}
\ea  
To simplify the notation, we introduce the following effective energy densities
\ba
w_f :=  k \; (\theta')^2 + b (\lambda -1)^2   -2 \mu  \lambda\cos \theta, \quad w_a :=  b(\lambda-1)^2 - 2 \lambda(w + \mu),
\ea 
that allow us to recast the energy functional in the more compact form
\ba
W = \int_{0}^{\bS} w_f \d S + \int_{\bS}^{L/2} {w_a} \d S + 2 \mu a.
\ea
Let us also introduce the varied quantities
\ba
\theta_\veps (S) := \theta(S) + \veps h(S), \quad \lambda_\veps (S) :=\lambda(S) + \veps u(S).
\ea
The variational procedure must explicitly include the fact that the end points $S = 0$ and $S = L/2$ are fixed, while the detachment point $S = \bS$ is not. We obtain 
\ba
\delta W = \int_0^\bS \left\{\left[ \frac{\pt w_f}{\pt \theta} - \left(\frac{\pt w_f}{\pt \theta'}\right)' \right] h + \frac{\pt w_f}{\pt \lambda} u\right\} \d S 
+  \int_\bS^{L/2}  \frac{\pt w_a}{\pt \lambda} u  \d S\nonumber\\
+ \left(\frac{\pt w_f}{\pt \theta'}\right)_{S=\bS^{-}} h(\bS^-) +
\left(w_f - w_a \right) _{S=\bS}  \delta \bS.
\label{variazione}
\ea
The first integral in Eq.\eqref{variazione} must vanish for any choice of $h(S)$ and $u(S)$. This leads to the equilibrium equations, valid in the buckled region $S \in [0,\bS)$,
\begin{align}
b(\lambda-1) = \mu \cos \theta, \label{compre} \\
k \theta'' -  \lambda \mu \sin \theta = 0. \label{eul}
\end{align}
The vanishing of the second integral of \eqref{variazione} for any $u(S)$ yields the equation that holds in the adhered region $S \in (\bS,L/2]$
\ba
b(\lambda -1)  = w + \mu.
\label{eq:ad}
\ea
From Eq. \eqref{eq:ad} we immediately deduce that stretch in the adhered region is constant: 
\ba
\lb = 1+\frac{w + \mu}{b}.
\label{eq:lb}
\ea
The presence of an adhesion $w$ introduces a line-tension force, and as a consequence the stretch $\lambda$ is not continuous at $S=\bS$. This is immediately apparent if we compare the horizontal forces at $S=\bS^{\pm}$, as given in Eq.\eqref{compre} (which holds for $S<\bS$) and in Eq.\eqref{eq:ad} (which holds for $S>\bS$).

We next discuss the boundary terms. We assume that $\theta_\veps (\bS+\veps \de \bS)$ is a regular function of $\veps$. Therefore, $\theta'(\bS^+)=0$ and $h(\bS^+)=0$ imply that $h(S^-)= - \theta'(\bS^-) \de \bS$ and the last two terms of \eqref{variazione} simplify to
\ba
\left(w_f - w_a - \frac{\pt w_f}{\pt \theta'} \right) \de \bS = 0.
\ea
Since this equation is requested to vanish for any $\de\bS$, we obtain the transversality condition
\ba
- k [\theta'(\bS)]^2 + 2 \lb w = 0, 
\label{eq:tra}
\ea
which relates the curvature of the beam at the detachment point with adhesion and stretching.

The Lagrange multiplier $\mu$ is determined by imposing the constraint
\ba
a = \int_{0}^{\bS} \lambda \cos \theta \d S + \lb\left(\frac{L}{2} - \bS\right) 
= \bx + \lb\left(\frac{L}{2} - \bS\right),
\label{eq:vinco}
\ea
where $\lb$ can be written in terms of $\mu$ according to Eq.\eqref{eq:lb}.

Finally, it is convenient, for the discussion of our results, to introduce the dimensionless parameter
\ba
\eps = \frac{L - 2a}{2a},
\label{eq:eps}
\ea
which measures the referential excess-length of sheet with the respect to the distance of the end-points. 

\section{Phase transition and critical threshold}
In this Section, we study the onset of a delamination blister. We show that the finite extensibility enable the delamination to occur at finite growth and above a critical threshold. Furthermore, we determine how the geometrical features of the blister depend on the geometry and the material parameters of the sheet.

\subsection{Numerical results and energy landscape}
\label{sec:2}

It is easy to find the analytical expressions of the completely adhered equilibrium solutions. As a matter of fact, in such cases, growth is accommodated by a pure compression of the sheet. These solutions are characterized by $\bS =0$, $\bx =0$, and, consequently, the only relevant equations are \eqref{eq:ad} and \eqref{eq:vinco}. These read
\ba
\lb_{\text{adh}} = \frac{1}{1+\eps}, \qquad \mu_{\text{adh}} = - w -  \frac{b \eps}{1+\eps}.
\label{eq:adh}
\ea
The stored energy for these configurations, $\Wa$, is immediately obtained by substituting the solution \eqref{eq:adh} into Eq.\eqref{eq:effective}
\ba
\Wa = k a \left(\frac{1}{\ell^2}\frac{\eps^2}{1+\eps} - \frac{2}{\ell_{ec}^2} \right).
\label{eq:energy_compressed}
\ea
By contrast, the energy associated with the buckled solution, $\Wb$, is not amenable for a simple analytical approximation and is best calculated with a numerical simulation. Figure 2 shows the comparison between the energies associated with each solution branch as a function of $\eps$, with $a=5$cm, $\ell_{ec}=1.35$cm and $\ell = 50\mu$m. The numerical values are consistent with those in Ref.\cite{Wagner:2013}. The energy of the adhered branch is given by Eq.\eqref{eq:energy_compressed}. By contrast, the collapsed solution comprises two branches. The upper branch, with higher energy, corresponds to smaller humps, whereas the lower branch corresponds to equilibrium solutions with larger humps. Finally, it is worth noticing that for sufficiently small values of $\eps$ only the adhered solution is admissible.
\begin{figure}[ht]
\centerline{\includegraphics[scale=1]{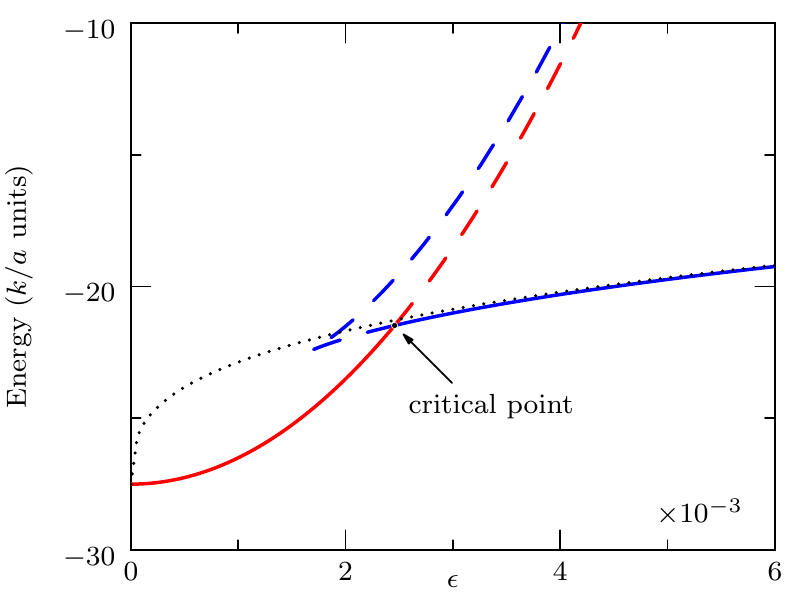}}
\caption{\label{figura_energie} Plot of the equilibrium free energy for the adhered solution (red) and the buckled solution (blue) with $a=5$cm, $\ell_{ec}=1.35$cm and $\ell = 50\mu$m. Solid lines represent solutions with least energy; dashed lines represent metastable states or maxima. For small $\eps$, only the adhered solution is admissible. The dotted line represents the energy of a perfectly inextensible strip as given in \eqref{eq:energy_buckled_incomp}. (Online version in colour.)}
\end{figure}
This energy landscape is reminiscent of a cusp bifurcation. Hence, with regards to the local stability of solutions, we immediately deduce that the upper-branch of the buckled solution is an energy maximum, while the remaining dashed lines are metastable states. A more rigorous analysis of the local stability confirms these predictions and has been carried out in a closely related problem with a similar bifurcation diagram in Ref\cite{napoli:2015}.

From the numerical simulations it emerges that, when the compression reaches a critical value, the beam buckles so to relax its internal stress and compression. Thus, the buckled solution is essentially determined by minimising the bending energy, while the compression energy can be neglected. It is then natural to assume that the solution with $\lambda=1$ correctly provides the buckled shape of the beam and its energy, also in the presence of a moderate adhesion potential. 
\section{Asymptotic analysis of the critical threshold}
\label{sec:3}

In the previous Section, we have seen that the critical excess-length at the transition can be obtained by comparing the energy of a purely compressed configuration with the energy of the buckled solution in the unstretchable case. To this end, we first report the analysis of the equilibrium solutions for an unstretchable rod. 

\subsection{Equilibrium of a growing sticky Elastica}
When $\lambda \equiv 1$, the buckled solution is determined only by Eq.\eqref{eul}. This can be written as a nonlinear pendulum equation
\ba
\theta'' + \tau \sin \theta =0,
\label{eq:pendolo}
\ea
where $\tau=-\mu/k$. A fist integral of Eq.\eqref{eq:pendolo} is
\ba
(\theta')^2 = 2 \tau (\cos \theta - \cos \theta_0),
\label{eq:primo_a}
\ea
where $\theta_0 := \theta(S_0)$ is the minimum value of $\theta$ in the range $[0,\bS]$. Thus, we can use the transversality condition \eqref{eq:tra} together with \eqref{eq:primo_a} to eliminate $\tau$ in favour of $\theta_0$
\begin{equation}
\tau = \ell_{ec}^{-2} \left(1-\cos\theta_0 \right)^{-1}.
\label{eq:tau}
\end{equation}
Thus, Eq.\eqref{eq:primo_a} reduces to
\ba
\theta' = \pm \ell_{ec} ^{-1}  \sqrt \frac{2(\cos \theta - \cos \theta_0)}{1-\cos \theta_0},
\label{eq:primo}
\ea
where the sign $-$ (respectively,  $+$) is to be used in the interval $S \in  (0, S_0)$ (respectively, $S \in (S_0 , \bS)$).  Eq.\eqref{eq:primo} is an ordinary differential equation that can be solved by separation of variables. The solution is 
\ba
\bS= -\frac{4 \ell_{ec}}{\sqrt 2 } \F[q_0],
\label{eq:asy1}
\ea
where $\F$ denotes the incomplete elliptic integral of first kind \cite{Abramowitz:1970} and, for ease of notation, we have set $q_0:= \{\theta_0/2, \csc(\theta_0/2)\}$

Similarly, we can exploit Eq.\eqref{eq:primo} to compute $\bar x$, the abscissa of the detachment point $\bar x := \int_0^\bS \cos \theta \d S$. With the help of Eqs.\eqref{eq:primo} and \eqref{eq:asy1}, we obtain
\ba
\bar x= -\frac{4 \ell_{ec}}{\sqrt 2}(1-\cos \theta_0) \E[q_0] + \bS \cos \theta_0,
\label{eq:asy2}
\ea
where $\E$ represents the incomplete elliptic integral of second kind \cite{Abramowitz:1970}. Finally, Eqs.\eqref{eq:vinco} and \eqref{eq:eps} yield the following identity between $\bS$ and $\bar x$
\ba
\eps = \frac{\bS - \bar x}{a}.
\label{eq:asy3}
\ea
The solutions of the nonlinear transcendental Eqs.(\ref{eq:asy1}-\ref{eq:asy3}) provide the values at equilibrium  of $\bS$, $\bar x$  and $\theta_0$. Hence, the total energy of the buckled configuration can be written in terms of these three quantities alone: 
\ba
\Wb = \frac{2 k}{\ell_{ec}^2}\left[ \frac{\bar x - \bS \cos \t0}{1- \cos \t0} - (1+\eps)a + \bS\right] .
\label{eq:energy_buckled_incomp}
\ea

\subsection{Asymptotic analysis}
We now provide an asymptotic expression for the critical threshold previously obtained numerically. To this end, we first look for the approximate expressions of $\bar x$, $\bS$ and $\t0$ as functions of $\eps$, by solving Eqs.(\ref{eq:asy1}--\ref{eq:asy3}). We use the strategy outlined in Ref.\cite{DePascalis:2014} to approximate the elliptic integrals as follows
\begin{align}
\F[q_0] & \approx \frac{\pi}{16}(9-\cos \t0) \sin \left(\frac{\t0}{2}\right), \\
\E[q_0] & \approx \frac{\pi}{64} (17-\cos \t0)\sin \left(\frac{\t0}{2}\right).
\end{align}
We look for solutions of Eqs. (\ref{eq:asy1}--\ref{eq:asy2}) in the form of a regular expansion in powers of $\eps^{1/3}$. Setting to zero the first three coefficients of the expanded equations, we obtain the following approximations
\begin{align}
\bar x & \approx \left(2 \pi ^2 a \eps  \ell_{ec}^2\right)^{1/3}
-\frac{7 a \eps}{8}, \label{eq:xb_asy}\\
\t0 & \approx -\left(\frac{4 \sqrt{2} a \eps}
{\pi  \ell_{ec}}\right)^{1/3}
-\frac{a \eps }{12 \sqrt{2} \pi  \ell_{ec}}.
\end{align}
The internal stress, $\tau$, is calculated from Eq.\eqref{eq:tau} and it is approximated by the following expression
\begin{equation}
\tau \approx \left(\frac{\pi}{2a \,\ell_{ec}^{2} \,\eps} \right)^{2/3} + \frac{1}{8\, \ell_{ec}^2},
\end{equation}
which manifestly shows that unstretchablility implies a divergent internal stress as $\eps \to 0$.

The elastic energy of the purely compressed solution and the energy of the incompressible buckled state, as given respectively in Eqs.\eqref{eq:energy_compressed} and \eqref{eq:energy_buckled_incomp}, are
\begin{align}
\Wa & \approx \frac{k a}{\ell_{ec}^2}\left(-2 + \frac{\ell_{ec}^2}{\ell^2}\eps^2\right) , \\
\Wb & \approx \frac{k a}{\ell_{ec}^2}\left(-2 + \left(54 \pi^2 \frac{\ell_{ec}^2}{a^2}\right)^{1/3} \eps^{1/3}- \frac{7}{4}\eps \right).
\end{align}
The critical threshold is determined by the condition $\Wa = \Wb$ which, to leading order, yields
\begin{equation}
\eps_{cr}\approx \left(\frac{54 \pi^2 \,\ell^{\,6}}{a^2 \ell_{ec}^4} \right)^{1/5}. 
\label{soglia_critica_a}
\end{equation}
This suggests the introduction of the dimensionless quantity \mbox{$\xi = \ell/(a\ell_{ec}^2)^{1/3}$}, so that \eqref{soglia_critica_a} simplifies to
\begin{equation}
\eps_{cr}\approx \left(54 \pi^2\right)^{1/5} \, \xi^{6/5}. 
\label{soglia_critica}
\end{equation}
The parameter $\xi$ controls how the solution approximately scales when we change the characteristic dimensions of the system: different materials and geometries show the same physical behaviour to leading order as long as they have the same value of $\xi$. 
\begin{figure}[ht]
\centerline{\includegraphics[scale=1]{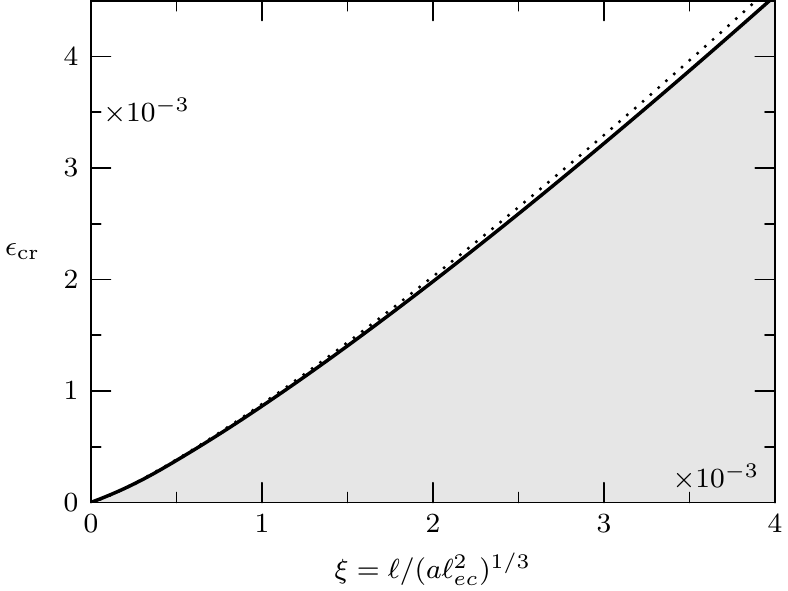}}
\caption{\label{fig:eps_critico} Critical threshold $\eps_{\text{cr}}$ as a function of the dimensionless parameter $\xi = \ell/(a \ell_{ec}^2)^{1/3}$. The solid line represents the numerical results, with $a=5$cm, $\ell_{ec}=1.35$cm and various film thickness from $0\mu$m to $\approx 139 \mu$m. The dotted line represents the analytical approximate result as given in Eq.\eqref{soglia_critica}. }
\end{figure}

\section{Conclusions}
\label{sec:4}
In this article we have presented numerical and analytical results for the shape and dimensions of delamination blisters. The presence of adhesion, a common feature in biology and in many microscopic contexts, is explicitly taken into account. In particular, we have considered the onset of delamination and we have shown that the compressibility is the key feature that allow us to study the phase transition and resolve the paradox of a diverging internal stress as $\eps \to 0$. An asymptotic analysis yielded simple expressions for the threshold and for the dimensions of the blisters, which are in excellent agreement with the numerical results. 

A comparison with a similar study in Ref.\cite{Wagner:2013}, where the deformations are not small, suggests that our results should be valid also beyond small deformations. The authors of \cite{Wagner:2013}, however, do not analyse the phase transition and the critical threshold as they assume that the sheet is inextensible. Therefore, they cannot determine the minimum blister size. By contrast, we can readily obtain $\bx_{\text{cr}}$ from the substitution of Eq.\eqref{soglia_critica_a} into Eq.\eqref{eq:xb_asy}:
\begin{equation}
\bx_{\text{cr}} = \left(12\pi^4 \,a\,\ell^2\,\ell_{ec}^2 \right)^{1/5}.
\end{equation}

Another closely related paper is Ref.\cite{Vella:2009}, where the authors consider a ``dual'' problem: the delamination of an \emph{inextensible} elastic strip adhering to an \emph{extensible} flat substrate. The comparison here is more indirect, but a careful identification and transcription of the key physical concepts shows that our results agree with those in Ref.\cite{Vella:2009}.




\vspace{0.5cm}
\begin{small}
\noindent\textbf{Conflict of Interest.} The authors declare that they have no conflict of interest.  
\end{small}

\end{document}